\documentclass[papersize,a4paper]{emulateapj}
\usepackage{apjfonts}
\usepackage{natbib}

\shorttitle{GRB Host Galaxies in Cosmological SPH Simulations}
\shortauthors{Niino et al.}

\newcommand{\Msun}{M_\odot}
\newcommand{\Muv}{M_{\rm UV}}
\newcommand{\Mstar}{M_\star}
\newcommand{\Zsun}{Z_\odot}
\newcommand{\Zcrit}{Z_{\rm crit}}
\newcommand{\Zgal}{Z_{\rm galaxy}}
\newcommand{\Pks}{P_{\rm KS}}
\newcommand{\Fgas}{f_{\rm gas}}

\begin{document}

\title
{Luminosity Distribution of Gamma-Ray Burst Host Galaxies at redshift $\lowercase{z}=1$ \\ 
in Cosmological Smoothed Particle Hydrodynamic Simulations: \\
Implications for the Metallicity Dependence of GRB\lowercase{s}}

\author{Yuu Niino \altaffilmark{1}, Jun-Hwan Choi \altaffilmark{2,3}, 
Masakazu A. R. Kobayashi \altaffilmark{4}, \\
Kentaro Nagamine \altaffilmark{2,5}, 
Tomonori Totani \altaffilmark{1} and Bing Zhang \altaffilmark{2}}

\email{niinou@kusastro.kyoto-u.ac.jp}

\altaffiltext{1}{Department of Astronomy, School of Science, Kyoto University,
Sakyo-ku, Kyoto 606-8502, Japan}
\altaffiltext{2}{Department of Physics and Astronomy, 
University of Nevada, Las Vegas, 
4505 S. Maryland Pkwy, Las Vegas, NV 89154-4002, U.S.A.}
\altaffiltext{3}{Department of Physics and Astronomy, University of Kentucky, Lexington, KY 40506-0055, U.S.A.}
\altaffiltext{4}{Optical and Infrared Astronomy Division, 
National Astronomical Observatory of Japan, 
Mitaka, Tokyo 181-8588, Japan}
\altaffiltext{5}{Visiting Researcher, Institute for the Physics and Mathematics 
of the Universe (IPMU), University of Tokyo, 5-1-5 Kashiwanoha, Kashiwa, 277-8583, Japan}

\begin{abstract}
We study the relationship between the metallicity of 
gamma-ray burst (GRB) progenitors and 
the probability distribution function (PDF) of GRB 
host galaxies as a function of luminosity  
using cosmological hydrodynamic simulations of galaxy formation. 
We impose a maximum limit to the gas metallicity in which GRBs can occur, 
and examine how the predicted luminosity 
PDF of GRB host galaxies changes in the simulation. 
We perform the Kolmogorov-Smirnov test, and show that the result from 
our simulation agrees with the observed luminosity PDF of core-collapse 
supernovae (SNe) host galaxies when we assume 
that the core-collapse SNe trace star formation.  
When we assume that GRBs occur only in a low-metallicity environment 
with $Z\lesssim 0.1 \Zsun$, GRBs occur in lower luminosity galaxies, 
and the simulated luminosity PDF becomes quantitatively consistent 
with the observed one.
The observational bias against the host galaxies of optically dark GRBs 
owing to dust extinction may be another reason for the lower luminosities 
of GRB host galaxies, but the observed luminosity PDF of GRB host galaxies 
cannot be reproduced solely by the dust bias in our simulation. 
\end{abstract}


\section{INTRODUCTION}
Long gamma-ray bursts (GRBs) are the brightest astronomical transient events, 
and they are important laboratories of high energy astrophysics 
in extreme conditions, as well as the tools to probe the 
high-redshift universe. 
For example, people have used GRBs to estimate the cosmic star formation 
rate (SFR) density at very high redshifts \citep[e.g.,][]{Kistler:09a}.
The observational association of some of the long GRBs with energetic Type 
Ic supernovae \citep[SNe, e.g.,][]{Hjorth:03a,Stanek:03a} is considered to be 
the evidence that at least some of the GRBs originate from  
core-collapses of very massive stars. 
\citet{Zhang:09a} applied multiple criteria to identify GRBs 
whose progenitors are related to core collapses of massive stars 
(the so-called "Type II" GRBs), and found that they mostly correspond 
to the traditional `long' GRB population. Hereafter `GRB' means 
Type II GRBs, and `SN' means core-collapse SN, unless otherwise stated.

Although GRBs may originate from the core-collapses of very massive stars, 
the occurrence rate of GRBs is much lower than that of normal SNe, 
and the conditions required for a GRB to occur from a SN still remain 
as one of the most outstanding questions in current astrophysics. 
A better understanding of required conditions for a GRB to occur 
would constrain the physical mechanisms of GRBs, 
and allow us to predict the GRB detection rate for 
the future observations more reliably. 

Some theoretical studies on the origin of GRBs 
using stellar evolution models suggest that a low metallicity environment may 
be a necessary condition for a GRB to occur \citep[e.g.,][]{MacFadyen:99a,Yoon:05a,Woosley:06a}. 
It has also been suggested from the observations 
that the metallicity distribution 
of GRB host galaxies at redshift $z < 0.25$ is significantly biased 
towards low metallicities compared to the expectation 
when GRBs are unbiased tracers of star formation \citep{Stanek:06a,Modjaz:08a}. 
However, reliable spectroscopic estimates of metallicities
are available only for galaxies at low redshifts \citep[$z \lesssim 0.5$;][]{Savaglio:09a}
while the majority of GRBs occur at higher redshift. 

Furthermore, some observations suggest that the GRB host galaxies are 
systematically fainter than those of the core-collapse SNe 
\citep[][hereafter F06]{Le-Floch:03a, Fruchter:06a}, indicating that the GRBs may preferentially occur 
in low metallicity environment, because fainter and lower mass 
galaxies generally have lower metallicities.  
These interpretations have also been supported by other theoretical studies 
using the models of galaxy formation and evolution \citep{Nuza:07a,Lapi:08a,Campisi:09a}. 

It is also reported that the GRB host galaxies at $z > 2$ 
have larger Ly$\alpha$ equivalent widths compared 
to general star-forming galaxies at similar redshifts \citep{Jakobsson:05a, Fynbo:02a, Fynbo:03a}, 
which is possibly a result of the stronger ionizing flux 
emitted from low-metallicity stellar population in GRB host galaxies \citep{Niino:09a}. 
Although the Ly$\alpha$ emission property of GRB host galaxies could 
be used as a metallicity indicator in the studies of GRB progenitors, 
our current understanding of Ly$\alpha$ transfer in the interstellar medium 
(ISM) is still inadequate to draw a robust conclusion \citep{Niino:09a}, 
and the samples of GRB host galaxies with Ly$\alpha$ detections is 
currently very small. 

The differences between the observed probability distribution functions (PDF) 
of GRB and SN host galaxies at $z \sim 1$ 
as a function of luminosity (hereafter `luminosity PDF') or size
is consistent with the hypothesis that GRBs occur preferentially in low-metallicity environment (F06). 
However, it is not straightforward to connect the difference 
in the luminosity and/or size of host galaxies 
to the metallicity difference of GRB and SN progenitors. 
In fact, some studies of GRB host properties claim 
that GRBs can be produced in higher metallicity environments 
than suggested by the stellar evolution models \citep{Wolf:07a,Kocevski:09a}. 
However, their results suffer from some uncertainties (see \S\,\ref{sec:previous}), and more tests are required to establish 
the connection between the metallicity dependence 
of GRBs and the luminosity PDF of GRB host galaxies. 

In this paper, we investigate the relationship
between the low metallicity preference of GRBs 
and the luminosity of GRB host galaxies at $z \sim 1$, 
where relatively large sample of observed GRB/SN host galaxies is available, 
using cosmological hydrodynamic simulations. 
We compare the predictions of our simulations 
with the observations to test if the low metallicity preference of GRBs 
predicted by the stellar evolution models is quantitatively consistent 
with the observed luminosity PDF of GRB host galaxies.
Several studies on this topic using galaxy formation models have already 
been carried out \citep{Nuza:07a,Lapi:08a,Campisi:09a}. 
However, quantitative comparisons between 
the predicted luminosity PDF of GRB host galaxies 
and the observed luminosity PDF have not been performed yet. 

This paper is organized as follows. 
In \S\,\ref{sec:simulation}, we briefly describe our simulation code 
and show some relevant properties of simulated galaxies, 
such as the luminosity function (LF), 
dust extinction, and luminosity--metallicity relationship. 
In \S\,\ref{sec:result}, we describe the modeling of GRB/SN event rate 
in the simulated galaxies, and compare the resulting luminosity PDF 
of the GRB/SN host galaxies to the observations. 
In \S\,\ref{sec:discussion}, we discuss the effect of possible 
observational bias by the optically dark GRBs on the luminosity PDF 
of GRB host galaxies. 
Then we compare our results with previous studies. 
Our conclusions are summarized in \S\,\ref{sec:conclusion}.

\section{NUMERICAL METHOD AND BASIC DATA}
\label{sec:simulation}
\subsection{Simulations and Galaxy Identification}
\label{sec:code}
We use the modified version of the tree-particle-mesh 
smoothed particle hydrodynamics (SPH) code 
GADGET-3 \citep[originally described in][]{Springel:05a}.
In this code, the SPH calculation is performed based 
on the entropy conservative formulation \citep{Springel:02a}.
Our conventional code includes radiative cooling by H, He, and 
metals \citep{Choi:09a}, heating by a uniform UV background of 
a modified \citet{Haardt:96a} spectrum, star formation, 
supernova feedback, phenomenological model for galactic winds, 
and a sub-resolution model of multiphase ISM and star formation 
\citep{Springel:03a}.

In this paper, we adopt the following fiducial cosmology which is consistent 
with the latest WMAP result: 
 $\Omega_m = 0.26$, $\Omega_{\Lambda} = 0.74$, 
 $\Omega_b = 0.044$, $h=0.72$, $n_{s}=0.96$, 
 and $\sigma_{8}=0.80$ \citep{Komatsu:09a, Komatsu:10a}.
We use two runs with different box sizes and resolution: N400L34 and N400L100 run.
See Table~\ref{table:sim} for the simulation parameters. 

We identify galaxies in the simulation at $z=1.0$ using 
a simplified variant of the SUBFIND algorithm \citep{Springel:01a,Choi:09a}. 
In more detail, the code first computes a smoothed baryonic density field 
to identify candidate galaxies with high density peaks. 
The full extent of these galaxies are found by adding gas and 
star particles to the galaxies in the order of declining density.
If all $N_{min}$ nearest neighbor particles have lower densities, 
this group of particles is considered as a new galaxy.  
Here, $N_{min}$ is the minimum number of gas and star 
particles that constitute one isolated galaxies.  
In this paper we set $N_{min} = 32$. 
If there is a denser neighbor, the particle is attached to the galaxy 
to which its nearest denser neighbor already belongs to. 
If two nearest neighbors belong to different galaxies 
and one of them has less than $N_{min}$ particles, 
then the two galaxies are merged.
If two nearest neighbors belong to different galaxies 
and both of them has more than $N_{min}$ particles, 
the particles are attached to the larger galaxy, 
leaving the other galaxy intact. 
In addition, the gas particles in galaxies should be denser 
than $0.01 \rho_{th}$, where $\rho_{th}$ 
is the star formation density threshold \citep{Springel:03a}.
In the current simulations, we use the SF threshold density of 
$n_{th}=0.6$\,cm$^{-3}$ \citep{Choi:09c}.

\begin{deluxetable}{cccccc}
\tablecolumns{6}
\tablewidth{0pc}
\tablecaption{Simulation Parameters}
\tablehead{
\colhead{Run} &  \colhead{Box-size} & \colhead{${N_{\rm p}}$} & 
\colhead{$m_{\rm DM}$} & \colhead{$m_{\rm gas}$} & \colhead{$\epsilon$} 
}
\startdata
N400L34    & 33.75  & $2\times 400^3$  & $3.60 \times 10^7$ & $7.33 \times 10^6$ &  3.375  \cr
\hline
N400L100   & 100.0 & $2\times 400^3$  & $9.12 \times 10^8$ & $1.91 \times 10^8$ &  6.45  
\enddata
\tablecomments{
Simulations employed in the present work. 
The box-size is given in units of $h^{-1}$Mpc, ${N_{\rm p}}$ is the particle
number of dark matter and gas (hence $\times\, 2$), $m_{\rm DM}$ and
$m_{\rm gas}$ are the masses of dark matter and gas particles in units of
$h^{-1} \Msun$, respectively, and $\epsilon$ is the comoving gravitational
softening length in units of $h^{-1}$kpc.  
The value of $\epsilon$ is a measure of spatial resolution. 
}
\label{table:sim}
\end{deluxetable}

\subsection{Distribution of Gas and Stars}
\label{sec:distribution}

\begin{figure}[t]
\begin{center}
\includegraphics[scale=0.68]{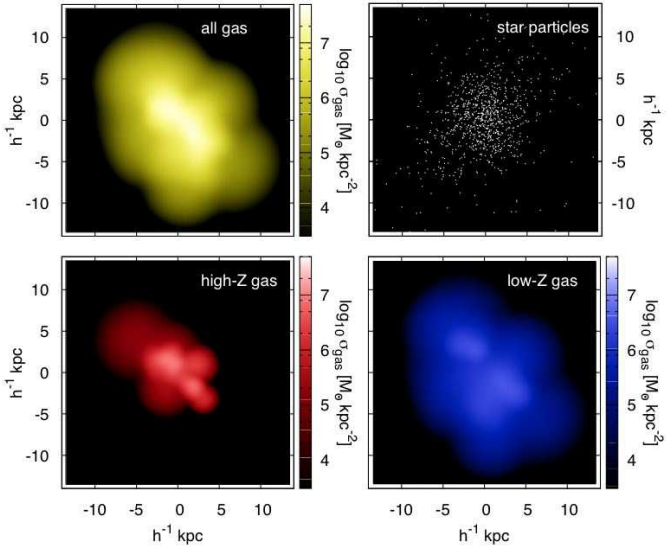}
\caption {
Distribution of gas ({\it top left}) and stars ({\it top right}) 
in an example galaxy from the N400L34 run.
This galaxy has $\Muv=-19.0$, $M_B=-19.3$, $\Mstar=3.5\times 10^{9} \Msun$, 
and $\Zgal = 0.33 \Zsun$. 
Distributions of gas with $Z>2.0\Zgal$ and $Z<0.2\Zgal$ 
are shown in the {\it bottom left} panel and the {\it bottom right} panel, respectively. 
}
\label{fig:example34}
\end{center}
\end{figure}

\begin{figure}[t]
\begin{center}
\includegraphics[scale=0.68]{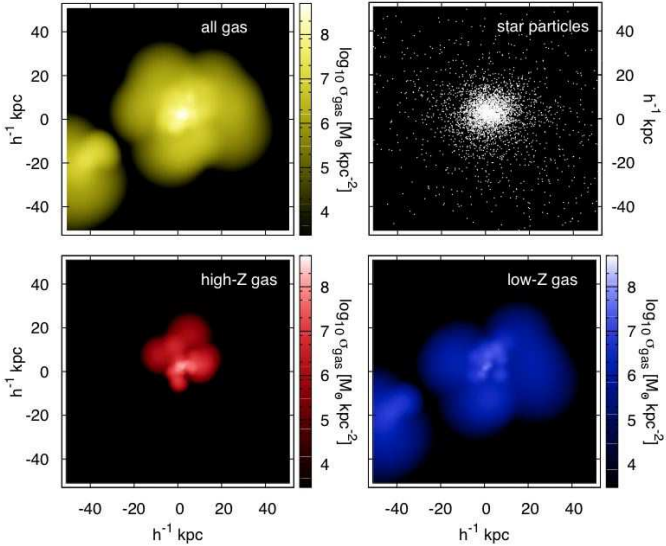}
\caption {
Same as Fig.\,\ref{fig:example34}, but for a more luminous galaxy.
This galaxy is from the N400L100 simulation, and has the 
following properties:
 $\Muv=-20.3$, $M_B=-21.8$, 
 $\Mstar=4.5\times 10^{11} \Msun$, and $\Zgal = 1.2\Zsun$. 
}
\label{fig:example100}
\end{center}
\end{figure}

We first examine the distribution of gas and stars in some simulated galaxies. 
Figure~\ref{fig:example34} is an example of a typical faint GRB host 
galaxy with $\Muv=-19.0$, 
while Figure~\ref{fig:example100} is a more luminous galaxy with $\Muv=-20.3$.
In the lower panels of Figures~\ref{fig:example34} and \ref{fig:example100}, 
the distribution of gas with high- and low- metallicity is shown separately. 
In both examples, the high-metallicity gas is 
concentrated at the center of its host galaxy, 
and the low-metallicity gas is distributed more broadly. 

It is reported that the observed GRBs are primarily located 
in the brightest regions of galaxies compared to SNe (F06), 
and the concentration of high-metallicity gas at the center of galaxies 
in our simulation may seem to contradict with 
the low-metallicity preference of GRBs.
However, if the GRBs occur in the young star clusters that are not 
heavily enriched yet \citep{Larsson:07a}, 
then their occurrence in bright regions would be unrelated to metallicity. 
We also see from this figure that our simulations do not have 
adequate resolution to resolve the disk structure with spiral arms, 
therefore it is difficult to discuss the spatial distribution of GRB sites
within a single galaxy using our current cosmological simulations.

\subsection{Dust Extinction Model}
\label{sec:modelEbv}

We compute the spectra and luminosities of simulated galaxies 
by applying the GALAXEV population synthesis code \citep{Bruzual:03a} 
to each constituent star particles 
and employing a simple model of dust extinction described below. 
To estimate the degree of extinction in each simulated galaxy, 
we assume that $E_{B-V}$ is proportional to the metal mass column density 
of the ISM: $\sigma_{\rm ISM} \times Z_{\rm ISM}$, 
where $\sigma_{\rm ISM}$ and $Z_{\rm ISM}$ are the gas column density 
and metallicity of each galaxy, respectively. 
The proportionality constant is fixed so that it agrees 
with the empirical estimate of gas-to-dust ratio in the Milky Way 
\citep[$N_{\rm H}/E_{B-V} = 5.8\times 10^{21}$\,cm$^{-2}$;][]{Bohlin:78a} when $Z_{\rm ISM} = \Zsun$:
\begin{equation}
E_{B-V} = \frac{f_{\rm H}}{5.8\times10^{21}} 
\frac{\sigma_{\rm ISM}}{m_p}  \frac{Z_{\rm ISM}}{\Zsun},
\end{equation}
where $f_{\rm H} = 0.75$ is the hydrogen mass fraction 
and $m_p$ is the proton mass. 
This formulation also agrees with the gas-to-dust ratio 
in the Small Magellanic Cloud (SMC) $N_{\rm H}/E_{B-V} \sim 10^{23}$ cm$^{-2}$ 
\citep[e.g.,][]{Bouchet:85a, Fitzpatrick:85a, Tumlinson:02a}, 
given that the metallicity in SMC is $\sim 10^{-1} \Zsun$. 
We use the \citet{Calzetti:97a} extinction curve 
to calculate the extinction at each wavelength.

To estimate $\sigma_{\rm ISM}$ for each simulated galaxy,
we assume that the ISM in each simulated galaxy follows 
the \citet{Hernquist:90a} mass profile with a total gas mass $M_{\rm ISM}$ of the galaxy. 
Then the mean $\sigma_{\rm ISM}$ within a radius $r$ is 
\begin{equation}
\sigma_{\rm ISM} = \frac{M_{\rm Hernquist}(< r)}{\pi r^2}
 = \frac{M_{\rm ISM}}{8\pi (r+r_c)^2},
\end{equation}
where the characteristic radius of the Hernquist mass profile $r_c$ 
is fixed such that $M_{\rm Hernquist}(< l_{\rm smooth,0}) = m_{\rm part,0}$. 
Here, $l_{\rm smooth,0}$ and $m_{\rm part,0}$ are the SPH smoothing length 
and the mass of the central gas particle with the highest gas density 
in the galaxy. 
For simplicity, we further assume $\sigma_{\rm ISM} = M_{\rm ISM} / 8\pi r_c^2$,  
which gives the value in the central region of the galaxy ($r \ll r_c$), 
and set $Z_{\rm ISM}$ to the metallicity of the central gas particle, 
because the highest density region of the galaxy is likely to be 
the main contributor to both the UV luminosity and GRB/SN production. 
Note that a single gas particle in our simulation represents 
a gas mass of $\sim 10^7 - 10^8 \Msun$, 
and $l_{\rm smooth,0}$ is typically on the order of kpc. 

\begin{figure}
\includegraphics[scale=0.75]{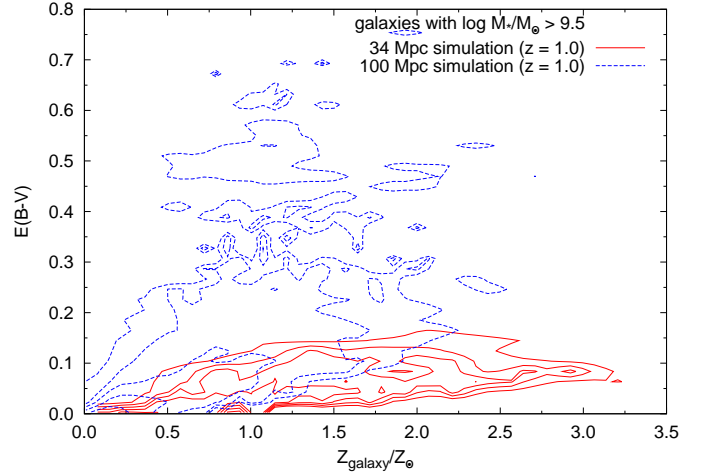}
\caption {
Extinction vs. metallicity for the simulated galaxies. 
The solid (red) and dashed (blue) contours are 
for the N400L34 and N400L100 runs, respectively. 
The contours represent $d^3n_{\rm gal}/dVdZdE_{B-V}$ = 0.01, 0.1,10, 100, 
and 1000 [$h^3$Mpc$^{-3}$mag$^{-1}$].}
\label{fig:Z-Ebv}
\end{figure}

In Figure~\ref{fig:Z-Ebv}, we show the distribution of simulated 
galaxies on the metallicity vs. extinction plane. 
Note that the metallicity used here is the mean gas metallicity 
of the whole galaxy ($\Zgal$). 
Only galaxies with $\Mstar > 10^{9.5} \Msun$ are shown in this figure, 
because the dust extinction in most of the lower mass 
galaxies are negligible ($E_{B-V} \lesssim 0.01$). 
The galaxies in the N400L34 run have lower $E_{B-V}$ than 
those in the N400L100 run for a given metallicity, 
indicating that $\sigma_{\rm ISM}$ is lower in the N400L34 run on average. 
This is consistent with the finding by \citet{Choi:09a} that 
the galaxies with a lower resolution run have higher gas mass fractions, 
because a higher resolution run can resolve higher density peaks 
and hence more gas is consumed for star formation.

\subsection{Luminosity Function of Simulated Galaxies}
\label{sec:luminosity}

\begin{figure*}
\begin{center}
\includegraphics[scale=0.8]{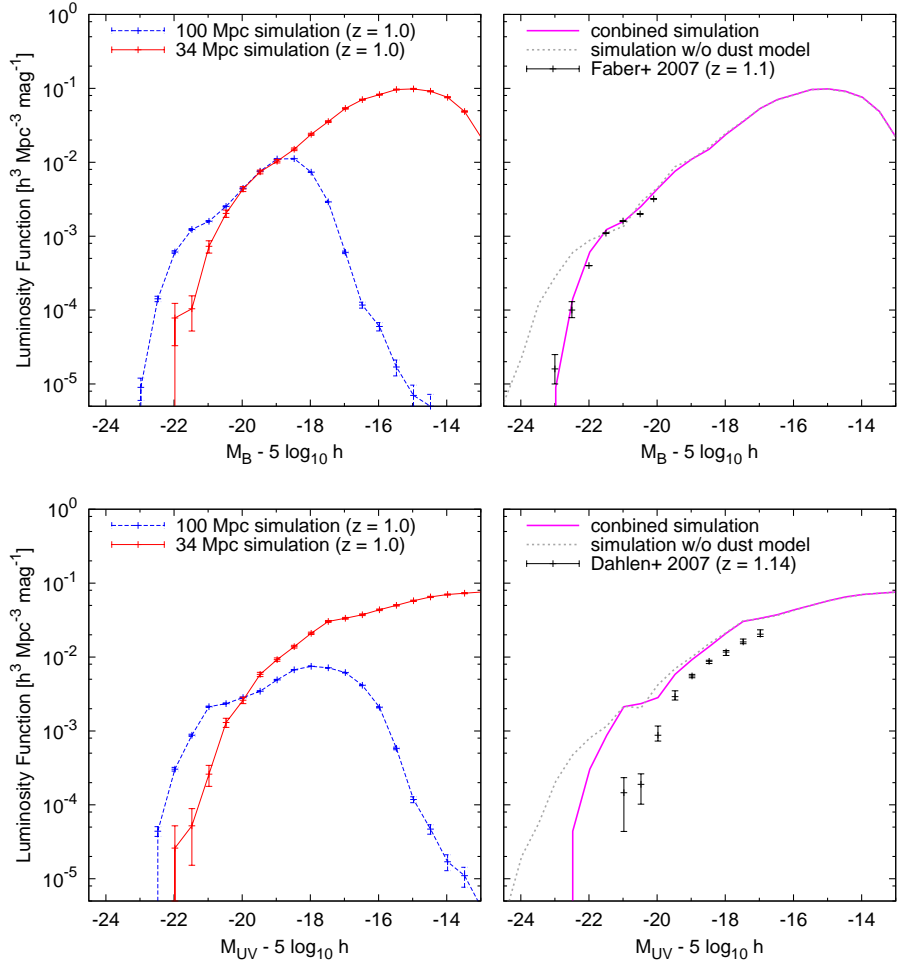}
\caption {
Luminosity functions of the simulated galaxies 
in the rest-frame $B$-band ({\it top panels}) 
and the rest-frame UV (2800\,\AA, {\it bottom panels}). 
{\it Top left panel}: The solid (red) and dashed (blue) lines 
are for the N400L34 and N400L100 runs, respectively.
The error bars show the Poisson errors. 
{\it Top right panel}: The solid (magenta) line represents 
the combined LF of both the N400L34 and N400L100 runs 
(i.e., the larger of the two simulation at each magnitude).
The LF without dust extinction (i.e., $E_{B-V} = 0$ for all galaxies) 
is shown with the dotted (grey) line.
The same line types are used in 
the {\it bottom panels} (rest-frame UV). 
The observed data points are taken from 
the DEEP2 survey for $B$-band \citep{Faber:07a},  
and the GOODS-S for the UV \citep{Dahlen:07a}. 
}
\label{fig:LF}
\end{center}
\end{figure*}

Figure~\ref{fig:LF} shows the LFs of simulated galaxies in the 
N400L100 and N400L34 runs, computed using the GALAXEV population 
synthesis code \citep{Bruzual:03a} and the dust extinction model 
described in \S\,\ref{sec:modelEbv}. 

Each simulation can resolve dark matter halos and galaxies only 
in a limited mass range due to its limited box-size and resolution.
In terms of the galaxy properties, the N400L100 run contains a 
larger number of massive, brighter galaxies than the N400L34 run 
due to its larger box size. The N400L34 run can resolve lower-mass, 
fainter galaxies better than the N400L100 run due to its higher resolution. 

The rest-frame $B$-band LFs are shown in the {\it top left} panel 
(dust extinction effect included) of Figure~\ref{fig:LF} with Poisson error bars. 
It is clear that the N400L100 run misses the faint galaxies 
with $M_B \gtrsim -19$, while the volume density of bright galaxies 
with $M_B \lesssim -20$ in the N400L34 run is smaller than that 
in the N400L100 run. 

Since the two runs probe galaxies with different masses and luminosities, 
we combine the two LFs by taking the larger of the two runs at each $M_B$, 
and compare the result with the observation by \citet{Faber:07a} 
in the {\it top right} panel of Figure~\ref{fig:LF}. 
The combined LF agrees quite well with the observation 
when the dust extinction effect is included. 

We also compare the rest-frame UV LF with the observation \citep{Dahlen:07a}
in the {\it bottom} panels of Figure~\ref{fig:LF}.
Contrary to the case of rest-frame $B$-band LF, 
the discrepancy between the observation and simulation is significant at 
the bright-end of the rest-frame UV LF.
We discuss the effects of this overprediction on our luminosity PDF 
of GRB host galaxies in \S\,\ref{sec:bright}.

Besides the bright-end overprediction, 
according to the comparison in Figure~\ref{fig:LF}, 
the simulation seems to overestimate the galaxy LF 
at any $\Muv$ by a factor of 1.5 at $z=1.0$.  
However, this over-estimation is probably within the range of uncertainties 
in our modeling and the cosmic variance in observations. 
It should be noted that the vertical shift of LF 
(e.g., the uniform overprediction at any $\Muv$) 
does not affect our luminosity PDF of GRB/SN host galaxies.

\subsection{Metallicity of Simulated Galaxies}
\label{sec:metal}

\begin{figure*}
\begin{center}
\includegraphics[scale=0.9]{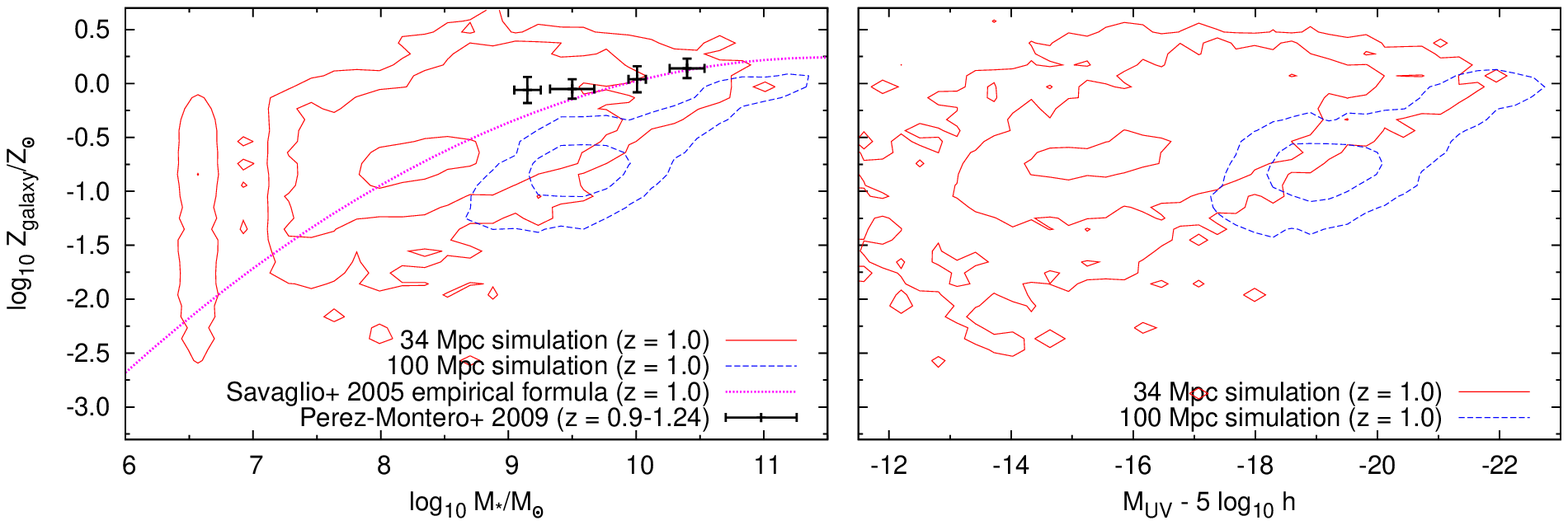}
\caption {
{\it Left:} Distribution of galaxies on $\Mstar - Z$ plane. 
The galaxies in the N400L34 and N400L100 runs are shown with 
the solid (red) and dashed (blue) contours, respectively.
The contours represent $d^3n_{\rm gal}/dVd($log$_{10}\Mstar)d($log$_{10}Z)$ 
= 0.005, 0.05, and 0.5 [$h^3$\,Mpc$^{-3}$].
The stellar mass and metallicity of the observed galaxies at 
$z \sim 1.0$ \citep{Perez-Montero:09a}
and the empirical formula of \citet{Savaglio:05a} for $z = 1.0$ 
galaxies are plotted together.
The empirical formula of \citet{Savaglio:05a} is based 
on the local and $z=0.7$ galaxies 
with a mass range of $\Mstar \sim 10^{8.5}$--$10^{11} \Msun$. 
{\it Right:} Same as the left panel, but on the $M_{\rm UV}$-$Z$ plane, 
where $M_{\rm UV}$ is the restframe UV magnitude.  
The contours represent $d^3n_{\rm gal}/dVd\Muv d($log$_{10}Z)$ 
= 0.002, 0.02, and 0.2 [$h^3$\,Mpc$^{-3}$\,mag$^{-1}$]. 
Colors are only for the online version. 
}
\label{fig:Zgalaxy}
\end{center}
\end{figure*}

The distribution of simulated galaxies on the 
stellar mass vs. metallicity ($\Mstar - Z$) plane is shown 
in the {\it left panel} of Figure~\ref{fig:Zgalaxy}. 
Here the metallicity means the mean gas metallicity of a galaxy. 
The N400L34 galaxies are distributed 
around the empirical formula at $z=1.0$ proposed by \citet{Savaglio:05a},  
which agrees well with the observational data points 
by \citet{Perez-Montero:09a}.
The N400L100 galaxies are distributed on the more massive and 
luminous side than the N400L34 galaxies as expected from the 
luminosity functions.
In both runs, more massive galaxies have higher metallicities
than lower mass galaxies. 

We note that the N400L100 galaxies have somewhat lower metallicities 
on average than those in the N400L34 run for the same $\Mstar$. 
This can be understood as follows. 
In our simulation, chemical enrichment is calculated 
as an instantaneous feedback from star formation, and 
the metallicity of a galaxy increases as the gas is turned into stars. 
Therefore the metallicity of a galaxy is anti-correlated with 
the gas mass fraction ($\Fgas$) of the galaxy. 
As mentioned in \S\,\ref{sec:modelEbv}, 
a higher resolution simulation can consume gas in star formation 
more efficiently than a lower resolution simulation. 
Furthermore, if we compare galaxies with similar $\Mstar$, 
galaxies with larger $\Fgas$ have larger total (star and gas) mass, 
and thus reside in lager simulation box. 
The N400L100 run has a higher $\Fgas$ compared to the N400L34 run
due to its larger box size and a lower resolution, leading to a lower metallicity of galaxies. 
As discussed in \S\,\ref{sec:luminosity}, 
the two simulation boxes complement each other, 
and each of the simulation cannot reproduce the observations alone
for a wide dynamic range.  
If we combine the two distributions of galaxies 
shown in the {\it left panel} of Figure~\ref{fig:Zgalaxy}, 
the overall distribution of the simulated galaxies agrees 
with the observations. 

It should be noted that the mean gas metallicity ($\Zgal$)
of a simulated galaxy depends on the threshold density 
of gas particles to be included in galaxies, 
which is a parameter in the galaxy finding algorithm 
described in \S\,\ref{sec:code}. 
If we include lower density gas particles than the current density threshold, 
each simulated galaxy may have lower $\Zgal$, 
because gas particles in the outskirts of a galaxy 
have lower metallicity than those near the center 
(see \S\,\ref{sec:distribution}). 
However, the dependence of $\Zgal$ on the grouping density threshold 
does not affect our luminosity PDF of GRB/SN host galaxies. 
Low density gas particles are not forming stars, 
and hence they do not contribute to the GRB production and UV luminosity. 

The distribution of the simulated galaxies on the $L_{\rm UV}-Z$ plane 
is also shown in the right panel of Figure~\ref{fig:Zgalaxy}, 
which is very similar to the $\Mstar-Z$ distribution.

\section{RESULTS}
\label{sec:result}
\subsection{Metallicity of GRB Sites}
\label{sec:GRBmetal}

In this section, we explore the metallicity of GRB sites. 
We first assume that the GRB occurrence rate in each simulated galaxy 
is proportional to the total SFR without a metallicity limit.
Then we examine the models in which the GRB rate is proportional 
to the total SFR in gas particles with $Z<\Zcrit$. 
Our interest is in the luminosity PDF of GRB/SN host galaxies, 
therefore we do not need to consider the absolute value of 
the event rate density.

\begin{figure}
\includegraphics[scale=0.72]{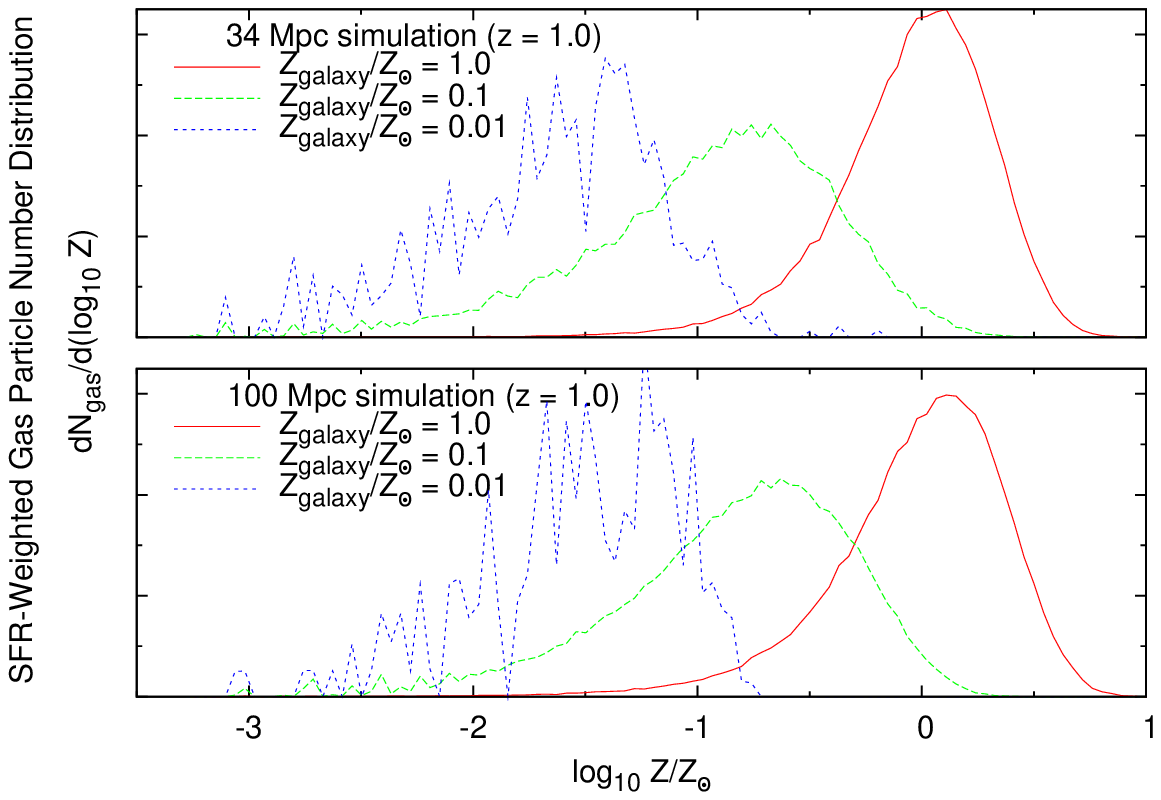}
\caption {
SFR-weighted metallicity distributions of gas particles in the 
simulated galaxies with 
log$_{10}(\Zgal/\Zsun) = 0.0\pm0.15,\ -1.0\pm0.15$, and $-2.0\pm0.15$ 
are plotted with solid (red), dashed (green), and dotted (blue) lines, 
respectively. 
In our model, we assume that the GRB rate is proportional to 
the sum of SFR for the gas particles with $Z<\Zcrit$.
The top and bottom panels are showing the distributions 
for the N400L34 and N400L100 simulations. 
The large fluctuation in the particle metallicity distribution 
in galaxies with $\Zgal = 0.01 \Zsun$ (blue dotted line) is 
due to the small number of gas particles included in 
such low-metallicity galaxies.
Colors are only for the online version. 
}
\label{fig:Zpart}
\end{figure}

In order to investigate the effect of metallicity limit, we first examine 
the SFR-weighted distribution of gas particles as a function of metallicity. 
In Figure~\ref{fig:Zpart}, we divide the galaxy population into 
three different samples according 
to the mean gas metallicity of the host galaxy ($\Zgal$). 
We can see that the metallicity of star forming gas particles 
has a wider range than that of $\Zgal$, 
showing the large dispersion of gas metallicity in each galaxy. 
We note that the high metallicity gas is concentrated 
in the high density regions of its host galaxy (see \S\,\ref{sec:distribution}) 
where SFR is high. Therefore the peak metallicity of SFR-weighted distribution is higher than the mean $\Zgal$ of the sample. 
The distributions for both N400L34 and N400L100 runs are very 
similar to each other. 

Assuming that the GRB rate is proportional to the sum of SFR 
in gas particles with $Z<\Zcrit$, 
we can see in Figure~\ref{fig:Zpart} that, if we set $\Zcrit = 0.1\,\Zsun$, 
then the GRB event rate in galaxies with $\Zgal \sim 1.0\,\Zsun$ (0.1\,$\Zsun$) 
would be reduced to a few percent (40\%) compared to the case of no $\Zcrit$. 
With $\Zcrit = 0.1\,\Zsun$, the event rate in galaxies with 
$\Zgal \sim 0.01\,\Zsun$ would be almost unchanged from the case 
of no $\Zcrit$. 
The large fluctuation in the distribution for galaxies with 
$\Zgal \sim 0.01\Zsun$ is due to the small number of 
gas particles included in such low-metallicity galaxies.

\subsection{Rest-frame UV Luminosity of the Host Galaxies}
\label{sec:host}

\begin{figure}
\includegraphics[scale=1.1]{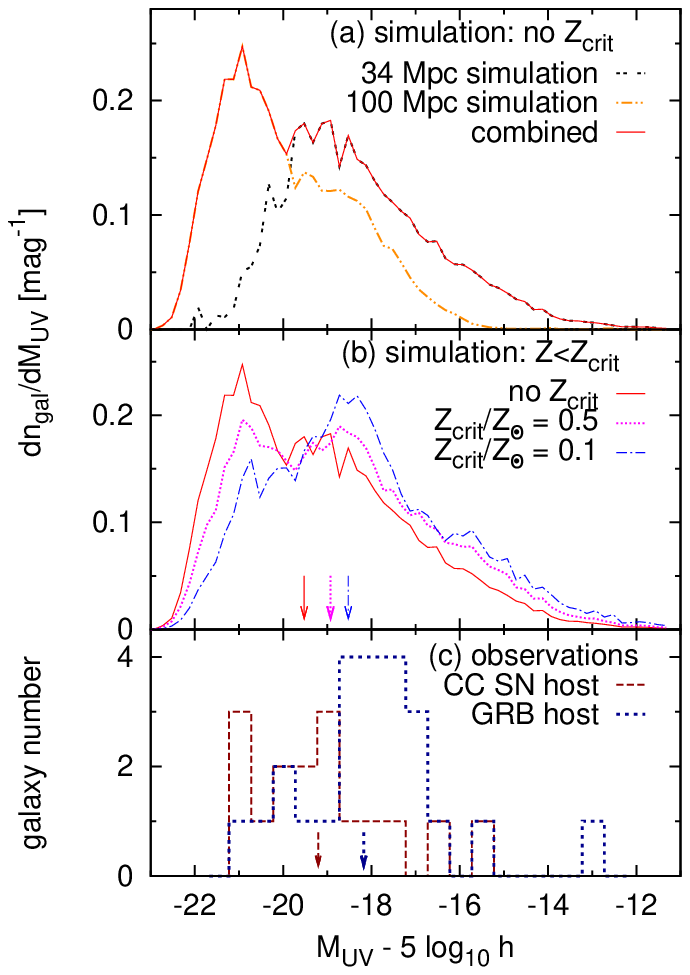}
\caption {
({\it a}) Luminosity PDF of the simulated host galaxies ($z=1.0$) 
without any metallicity cutoff. 
The results of the N400L34 and N400L100 run are 
plotted with double-dotted (black) 
and double-dot dashed line (orange), respectively. 
The combined distribution of the two simulations 
(i.e., the larger of the two distributions) 
is plotted with solid line (red).
({\it b}) Combined luminosity PDFs of the simulated host galaxies 
with no $\Zcrit$, $\Zcrit/\Zsun = 0.5$, and 0.1 
are plotted with the solid (red), dotted (magenta), 
and dot-dashed line (blue), respectively. 
({\it c}) Luminosity PDFs of the observed sample 
of GRB host galaxies and SN host galaxies at $z < 1.2$ (F06) 
are plotted with dotted (dark blue) and dashed (dark red) histograms, 
respectively. The arrows indicate the median of the distributions. 
Colors are only for online version. 
}
\label{fig:hostLF}
\end{figure}

Next, in order to discuss the luminosity PDF of host galaxies, 
we consider the SFR-weighted fraction of galaxies as a function of 
rest-frame UV luminosity as shown in Figure~\ref{fig:hostLF}.
The luminosity PDF in the case of no $\Zcrit$ is plotted 
in {\it panel (a)} for the N400L34 and N400L100 runs. 
As mentioned in \S\,\ref{sec:luminosity}, the two simulations have 
different resolving power of galaxies with different mass scales, 
therefore we combine the two distributions by taking the larger 
of the two runs in units of $\Msun$\,yr$^{-1}\,h^3$\,Mpc$^{-3}$\,mag$^{-1}$ 
as shown in {\it panel (a)}. 
The distributions are normalized so that the integration of 
the combined luminosity PDF for each $\Zcrit$ is unity. 
Hereafter we discuss the combined luminosity PDFs unless otherwise mentioned.

In {\it panel (b)} of Figure~\ref{fig:hostLF}, 
the three combined luminosity PDFs of the simulated host galaxies 
with $\Zcrit=$ none, 0.5$\Zsun$, and 0.1$\Zsun$ are shown. 
The luminosity PDF shifts to the fainter magnitudes as 
$\Zcrit$ becomes lower. 
In the case of $\Zcrit=0.1\Zsun$, the host galaxies 
are fainter than in the no-$\Zcrit$ case by $\sim 1$ magnitude.
This result is consistent with the luminosity PDFs of 
the observed GRB/SN host galaxies at $z<1.2$ (the F06 sample)
as shown in {\it panel (c)}. 
The redshift criterion of $z<1.2$ was proposed in F06 
to reduce the effect of galaxy evolution on 
the comparison between the GRB and SN host galaxies. 
The luminosity PDF of F06 was in the observed frame $V$-band, 
which is close to the UV in the galaxy's rest-frame. 
The median values of each distribution in {\it panels (b)} and {\it (c)}
are indicated by the arrows. 

It should be noted that, using a larger observed sample, 
\citet{Svensson:10a} found a smaller difference between 
the GRB and SN host galaxy luminosity PDFs than that claimed by F06. 
However, it is difficult to understand the selection effect 
in the GRB host sample of \citet{Svensson:10a}, 
because a part of it is drawn from the GHostS project database, 
which consists of publicly available data of GRB host observations 
done by various groups with differing instruments \citep{Savaglio:09a}. 
On the other hand, the F06 sample is a complete sample 
of host galaxies for all 42 GRBs that have optical afterglow 
detections ($>3\sigma$), 
therefore the sampling effect would be smaller 
in the F06 sample than in the \citet{Svensson:10a} sample.
Thus we use the GRB host galaxy sample by F06 
as a reference data for the discussions in this paper. 
The 24 out of the 42 GRBs in the F06 sample are at $z<1.2$. 
We note that the SN host galaxies in the \citet{Svensson:10a} sample 
is significantly fainter than those in the F06 sample, 
and this is probably one of the causes for the smaller difference 
between the GRB and SN host galaxies. 
But the reason for the difference between the SN host samples 
of F06 and \citet{Svensson:10a} is not known. 

Even the F06 sample would suffer from 
the observational bias against dusty host galaxies. 
We will discuss the selection effect 
caused by the dust extinction in \S\,\ref{sec:dustbias}. 
We note that the luminosity PDFs studied by F06 and 
\citet{Svensson:10a} are at different wavelengths 
[rest-frame UV for F06 versus rest-frame $B$ \& $V$-band for \citet{Svensson:10a}], 
though this cannot explain the difference in their results. 
The difference in the luminosity PDFs of GRB and SN host galaxies 
should be larger in the optical than in UV, 
given that many of observed GRB host galaxies are 
extremely blue \citep[e.g.,][]{Le-Floch:03a}. 

\subsection{Cumulative Luminosity PDF}
\label{sec:cumulative}

\begin{figure}
\includegraphics[scale=0.72]{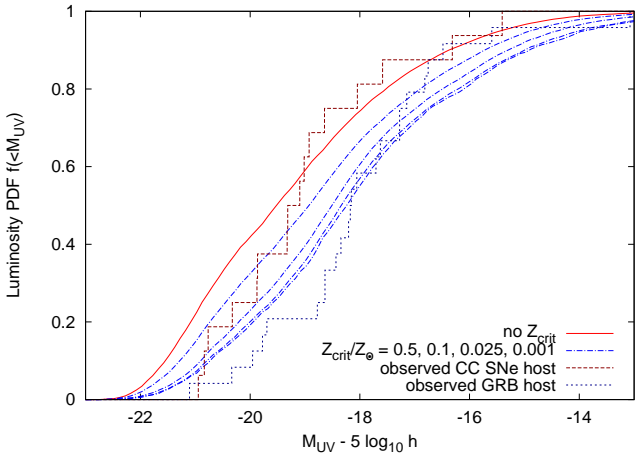}
\caption {
Cumulative luminosity PDFs of the simulated host galaxies 
is shown together with those of the observed GRB/SN host galaxies 
of F06 at $z < 1.2$, based on the histograms shown in Fig. \ref{fig:hostLF}.
The PDF without a metallicity cutoff (no $\Zcrit$) is shown 
with the solid line (red), and the those with $\Zcrit/\Zsun=$ 
0.5, 0.1, 0.025, and 0.001 are represented by the dot-dashed lines (blue) 
from top to bottom, respectively. Colors are only for online version. 
}
\label{fig:cumUV}
\end{figure}

We plot the cumulative luminosity PDFs of 
the simulated and observed host galaxies in Figure~\ref{fig:cumUV}. 
The simulated PDFs are shown for the cases of 
$\Zcrit/\Zsun =$ none, 0.5, 0.1, 0.025, \& 0.001 (top to bottom). 
The observed host galaxy sample at $z<1.2$ of F06 is plotted together
as histograms. 
The shift of luminosity PDF to fainter magnitudes 
is clearly seen also in the cumulative plot, 
reproducing the difference in the observed GRB and SN-host 
galaxy luminosity PDFs. 
However, the distribution for no-$\Zcrit$ case overpredicts 
the bright-end of the observed SN-host luminosity PDFs, 
and the $\Zcrit \leq 0.1\Zsun$ case overpredicts 
the bright-end of the observed distribution of GRB host galaxies. 
This is expected from the overprediction of the UV LF 
at the bright-end as we discussed in \S\,\ref{sec:luminosity} 
(see the bottom panel of Fig. \ref{fig:LF}). 
We discuss the effects of this overprediction on our conclusion 
in \S\,\ref{sec:bright}.

\subsection{Kolmogorov-Smirnov Test}
\label{sec:KStest}

\begin{figure}
\includegraphics[scale=0.7]{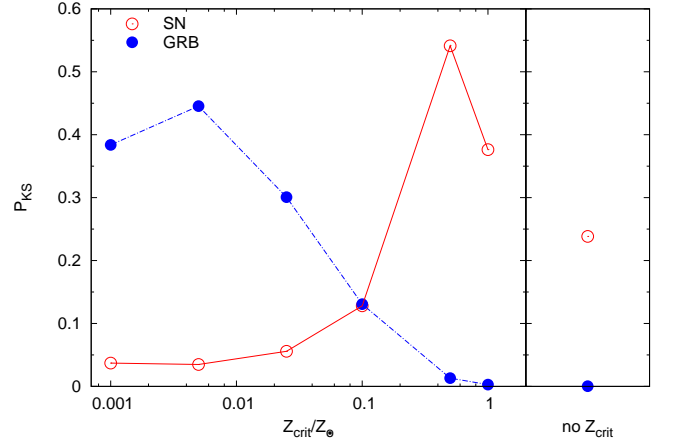}
\caption {
K-S probabilities (i.e., 1.0 minus the rejection confidence level for 
consistency) of the luminosity PDFs of the observed 
and the simulated host galaxies is plotted against $\Zcrit$.  
The symbols are for the SN host galaxies (red empty circles) and 
GRB host galaxies (blue filled circles). 
}
\label{fig:KStest}
\end{figure}

To examine the consistency between the simulated and observed 
cumulative luminosity PDFs shown in Figure~\ref{fig:cumUV}, 
we employ the one-sample Kolmogorov-Smirnov (K-S) test, 
which calculates the probability that a sample distribution 
is consistent with a given PDF. 
Here, the sample is the observed luminosity of GRB/SN host galaxies, 
and the PDFs are the results of our simulation. 
Figure~\ref{fig:KStest} shows the K-S probability 
$\Pks$ (i.e., one minus the rejection confidence level for consistency) 
as a function of  $\Zcrit$ of simulated host galaxies. 
The $\Pks$ takes a value between 0 and 1, 
and $\Pks<10^{-4}$ means that the observed sample is dissimilar to 
the simulated one at a $3\sigma$ level, 
while a larger $\Pks$ means a higher probability that 
the observed sample is derived from a PDF calculated in our simulation. 
For the SN host galaxies, we find $\Pks > 0.05$ when 
$\Zcrit \geq 0.03\Zsun$, and $\Pks > 0.2$ when $\Zcrit \geq 0.5\Zsun$. 
On the other hand, for the GRB host galaxies, 
we find $\Pks > 0.05$ (0.2) when $\Zcrit/\Zsun \leq 0.1$ (0.03).
The KS probabilities are the highest when $\Zcrit/\Zsun=0.005$ 
and 0.5 for the GRB and SN host galaxies, respectively. 
This clearly supports the low-metallicity preference for GRB host galaxies. 

The luminosity PDF of the simulated host galaxies does not change 
significantly for $\Zcrit \leq 0.1\Zsun$, 
and the agreement with the observed PDF is reasonably good 
for all $\Zcrit$ values for $\Zcrit < 0.1\Zsun$.
Therefore we cannot put a lower limit to $\Zcrit$ based on this comparison. 
However our simulation cannot resolve the galaxies with $\Mstar<10^8\Msun$, 
therefore it is possible that the above result suffers from 
the resolution limit of our simulation. 
We should be cautious about the results with $\Zcrit < 0.1\Zsun$, 
but our simulation contains galaxies as faint as $\Muv \sim -14$
(see the bottom panels of Figure~\ref{fig:LF}), 
which are well below the typical luminosity of 
the simulated host galaxies with $\Zcrit < 0.1\Zsun$.

\subsection{Overprediction of UV-bright Galaxies}
\label{sec:bright}

In \S~\ref{sec:luminosity} and Figure~\ref{fig:LF}, 
we showed that our N400L100 simulation overpredicts 
the bright-end of the rest-frame UV LF 
even after the dust extinction correction.  
We note however that our simulation agrees 
very well with the observed rest-frame $B$-band LF, 
which suggests that either our simulation overpredicts 
the amount of young stars compared to the older stellar population, 
or the two sets of observations are inconsistent with each other, 
or else the Calzetti extinction curve that we adopted 
is not appropriate for the observed galaxies.
The source of this discrepancy is unclear, therefore we focus 
only on the rest-frame UV LF in this subsection. 

To quantify the effect of the bright-end overprediction on our 
luminosity PDFs, we perform the same K-S test for 
the following two different test models that modify the shape of 
the rest-frame UV LF by hand.  

\vspace{0.2cm}
 $\bullet$ {\bf Model 1}:
We double the gradient of the rest-frame UV LF at the bright-end
($\Muv<-20$) as follows: 
\begin{eqnarray}
\begin{array}{ll}
	M_{\rm UV,model\ 1} = \Muv + \Delta\Muv \\
	\Delta\Muv = \left\{
	\begin{array}{ll}
		-0.5\times (\Muv + 20), & \ \ \Muv < -20 \\
		0, & \ \ \Muv \geq -20. \\
	\end{array} \right.
\end{array} \label{eq:modulation}
\end{eqnarray}

The SFR in galaxies with $\Muv<-20$ is also modified in a consistent manner, 
motivated by the inference that the overprediction of the bright-end 
of the UV LF is caused by an overestimation of SFR in massive galaxies 
\citep[e.g., due to lack of active galactic nucleus feedback in our simulation;][]{Choi:09c}. 
The change of SFR in each simulated galaxy is as follows:
\begin{equation}
\Delta({\rm log_{10}SFR}) = \Delta({\rm log_{10}}L_{\rm UV}) = -\frac{\Delta\Muv}{2.5}.
\label{eq:dSFR}
\end{equation}

\vspace{0.2cm}
 $\bullet$ {\bf Model 2}:
We shift the entire UV LF uniformly by $\Delta\Muv = 1.0$ to fit the observation. 
In this model, the luminosity PDF will also shift by the same amount of $\Delta\Muv$, without changing its shape.

\begin{figure}
\includegraphics[scale=1.0]{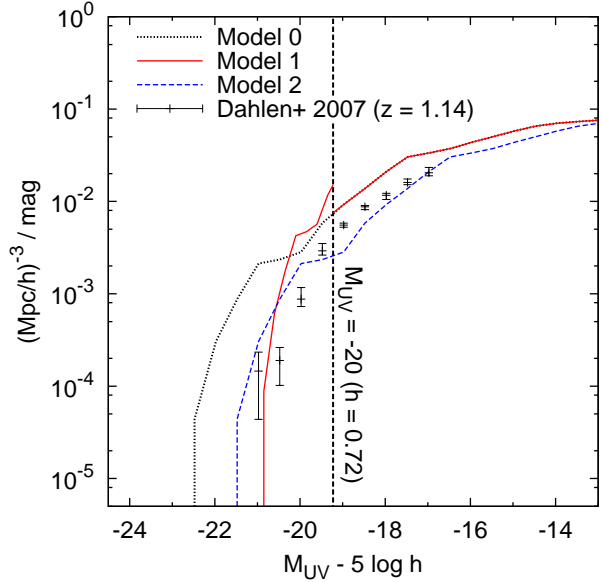}
\caption {
Rest-frame UV luminosity functions of the simulated galaxies in Model 0, 1 \& 2 
are plotted together with the observation \citep{Dahlen:07a}. 
The dust extinction model is included. 
The original result of our simulation 
(Model 0, identical to the combined simulation shown 
in the {\it bottom right panel} of Fig.\,\ref{fig:LF}) 
is shown with the dotted (black) line. 
The modified LF with the doubled gradient of the bright-end slope (Model 1) 
and the horizontally shifted LF by $\Delta\Muv = 1.0$ (Model 2) 
are represented by solid (red) and dashed (blue) curves, respectively. 
Vertical dashed line represents $\Muv = -20$, 
which is the threshold of the bright-end modification in Model 1. 
}
\label{fig:modLF}
\end{figure}

\vspace{0.2cm}
Both of these models are ad hoc modifications to fit the observed 
data points, but they would allow us to evaluate the effects of 
overprediction of UV LF by our simulation.  
For comparison purposes, we call the earlier calculations presented 
in \S\,\ref{sec:cumulative} -- \ref{sec:KStest} the {\bf Model 0}. 
We show the LFs of Model 0, 1 \& 2 galaxies in Figure ~\ref{fig:modLF}. 
One sees that the overprediction of the bright-end 
in Model 0 is removed in Model 1 \& 2. 
The uniform overprediction at any $\Muv$ is not removed in Model 1, 
but the uniform overprediction does not affect 
the luminosity PDF of GRB/SN host galaxies, 
as we have mentioned in \S~\ref{sec:luminosity}.

\begin{figure}
\includegraphics[scale=0.7]{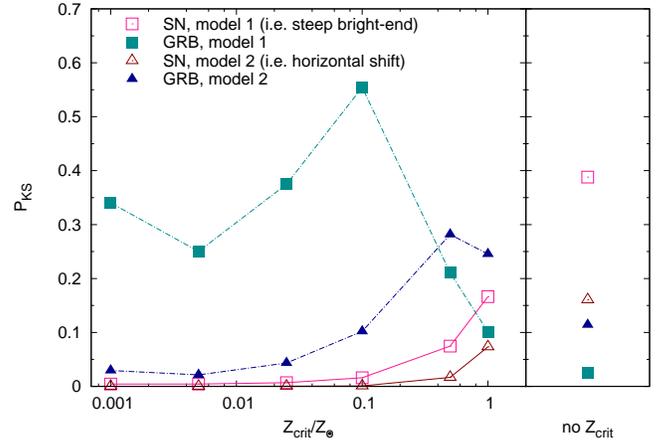}
\caption {
Same as Fig.\,\ref{fig:KStest}, 
but for the two different models of modified UV LF (Model 1 \& 2). 
The Model 1 \& 2 are represented by the squares and triangles, respectively. 
}
\label{fig:modKS}
\end{figure}

Figure~\ref{fig:modKS} shows the K-S probabilities 
of Models 1 \& 2 for the observed GRB/SN host galaxies. 
For the SN host galaxies, the no-$\Zcrit$ case has the highest $\Pks$, 
and $\Pks$ decreases with decreasing $\Zcrit$. 
For the GRB host galaxies, the Model 1 gives the highest probability 
of $\Pks \sim 0.6$ at $\Zcrit=0.1\,\Zsun$, 
and $\Pks \approx 0.3-0.4$ at $\Zcrit <  0.1\,\Zsun$.  
In the case of Model 2, 
the trend is opposite of what we saw in Figure~\ref{fig:KStest} 
for $\Zcrit<0.5\,\Zsun$ in the sense that $\Pks$ increases with 
increasing $\Zcrit$.
But the Model 2 still prefers $\Zcrit = 0.5\Zsun$ to the no-$\Zcrit$ case. 
The Model 2 gives $\Pks < 0.2$ across all $\Zcrit$ values 
for the SN host galaxies, which means that it doesn't agree well 
with the observations of SN host galaxies compared to the Models 0 and 1.

\section{Discussions}
\label{sec:discussion}

\subsection{Effect of Dust Extinction in the Host Galaxies}
\subsubsection{Observational bias against dusty host galaxies}
\label{sec:dustbias}

In this section we discuss the possibility that the systematic faintness 
and the low-metallicity of the observed GRB host galaxies is caused 
by the observational bias against dusty host galaxies, 
{\it not} by the intrinsic nature of GRB progenitors. 
It is difficult to identify the GRB host galaxy without a detection 
of an optical afterglow, and the afterglow flux may be reduced 
significantly by the dust extinction effect.
In fact, some GRBs have optical to X-ray afterglow spectral 
index $\beta_{OX} < 0.5$\footnote{The optical to X-ray 
afterglow spectral index $\beta_{OX}$ is defined 
as $f_{\rm opt} / f_X = (\lambda_{\rm opt} / \lambda_X) ^{\beta_{OX}}$, 
where $f_{\rm opt}$ and $f_X$ are the optical and X-ray fluxes, respectively. }, 
which is smaller than the expectation from the standard 
external shock model of GRB afterglows \citep[e.g.,][]{Jakobsson:04a}. 
We note that a part of the optically dark GRB could also be caused 
by the effects other than the dust extinction in their host galaxies, 
such as the attenuation by neutral hydrogen in the 
intergalactic medium at $z\gtrsim 6$ \citep[e.g.,][]{Totani:06a, Nagamine:08a}.

\begin{figure}
\includegraphics[scale=0.7]{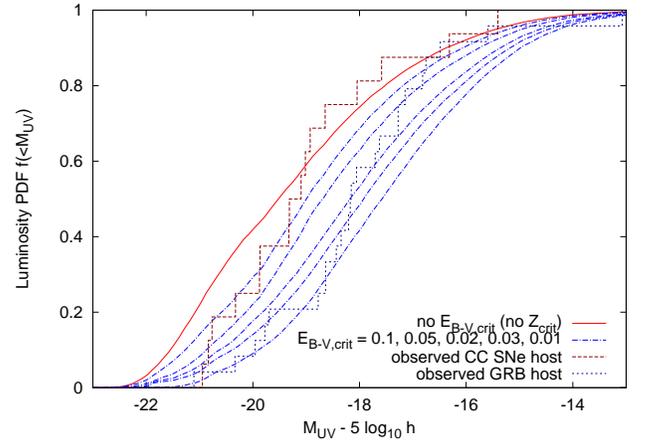}
\caption {
Same as Fig.\,\ref{fig:cumUV}, 
but here the galaxy is excluded from the sample 
if the average extinction value of galaxy exceeds $E_{B-V, {\rm crit}}$. 
We use Model 0 in this figure.
}
\label{fig:EcumUV}
\end{figure}

In order to examine the effect of dust extinction in the host galaxy 
(as opposed to the extinction in the immediate nearby environment of the GRB), 
here we exclude the galaxy from the sample 
of simulated host galaxies if $E_{B-V}\geq E_{B-V, {\rm crit}}$, 
where $E_{B-V}$ is the average extinction of each galaxy 
computed in \S\,\ref{sec:modelEbv}. 
The GRB event rate in each galaxy is assumed 
to be proportional to the SFR in the galaxy. 
Figure~\ref{fig:EcumUV} shows the cumulative luminosity PDF 
of simulated host galaxies with various $E_{B-V, {\rm crit}}$. 
The no-$E_{B-V, {\rm crit}}$ case ({\it red solid} line) represents 
the luminosity PDF of GRB host galaxies when the GRB rate simply traces 
the SFR, and it is identical to the no-$\Zcrit$ case in Figure~\ref{fig:cumUV}. 
The luminosity PDF of the simulated host galaxies shifts to 
the fainter magnitudes with smaller $E_{B-V, {\rm crit}}$, 
similarly to the $\Zcrit$ dependence shown in Figure~\ref{fig:cumUV}.

\begin{figure}
\includegraphics[scale=0.7]{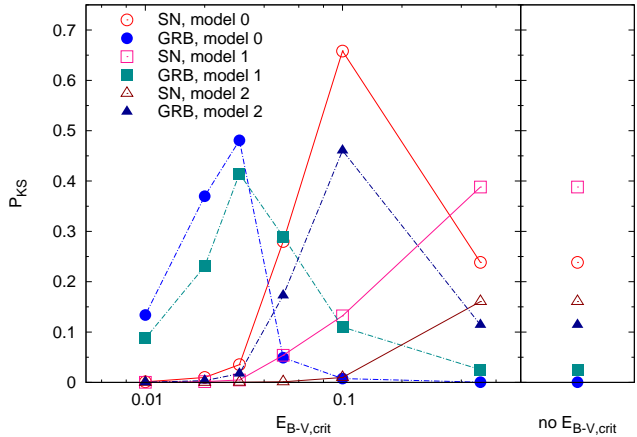}
\caption {
Same as Figs.\,\ref{fig:KStest} and \ref{fig:modKS}, 
but here $\Pks$ is plotted against $E_{B-V, {\rm crit}}$ instead of $\Zcrit$.
}
\label{fig:EKS}
\end{figure}

Figure~\ref{fig:EKS} shows the K-S probability versus $E_{B-V, {\rm crit}}$ 
for the luminosity PDF of the Model 0 galaxies given in 
Figure~\ref{fig:EcumUV}, together with the results of Model 1 \& 2 galaxies. 
For Model 0 \& 1, a reasonable agreement ($\Pks > 0.2$) between 
the simulated and observed distributions of the SN host galaxies 
is found for a wide range of critical extinction ($E_{B-V, {\rm crit}} \geq 0.05$), 
while for the GRB host galaxies 
the same level of agreement is achieved only when $E_{B-V, {\rm crit}} \leq 0.03$. 

The above requirement of $E_{B-V, {\rm crit}} \leq 0.03$ for the GRB host galaxies 
of the Models 0 \& 1 does not seem to agree with observations. 
Some follow up observations of optically dark GRBs suggest that 
a typical dust extinction to make a GRB optically dark is 
$A_V \sim 1$ \citep[e.g.,][$E_{B-V} \sim 0.3$]{Perley:09a}, 
which is much larger than the above requirement of $E_{B-V, {\rm crit}}<0.03$.
Such large values of critical extinction ($E_{B-V, {\rm crit}} > 0.1$) do not 
significantly change the luminosity PDF of simulated host galaxies. 
A small dust extinction such as $E_{B-V} < 0.03$ is unlikely 
to have a significant impact on the GRB optical afterglow observations. 

For the Model 2, 
the agreement with the observed GRB hosts 
is good ($\Pks = 0.48$) at $E_{B-V, {\rm crit}}=0.1$, 
however, $\Pks > 0.2$ is not obtained for the SN host galaxies 
with any values of $E_{B-V, {\rm crit}}$. 
The SFR-weighted luminosity PDF of the Model 2 galaxies are 
too faint to agree with the observation of SN host galaxies 
even in the no-$E_{B-V, {\rm crit}}$ case. 
Hence the difference of luminosity PDFs between the GRB host 
and SN host galaxies cannot be reproduced by the Model 2. 

A caveat here is that the $E_{B-V}$ is computed in our model 
assuming simplified structure of ISM in each galaxy 
without considering smaller scale structures such as molecular clouds. 
Though our model of dust extinction reproduces the 
observed $B$-band LF of field galaxies, 
the model may be not representative of dust extinction 
for lines of sight to GRBs, 
if small scale structures near GRBs play an important role 
in the extinction of GRB optical afterglows. 
Recent observations of optically dark GRBs actually suggest that 
the effect of dust on small scales may be important
\citep{Rol:07a,Castro-Tirado:07a,Perley:09a,Hashimoto:10a,Holland:10a}. 
If the typical extinction in lines of sight to GRBs is 
greater than the average extinction in its host galaxy 
by one order of magnitude but yet proportional to the average extinction,  
the luminosity PDF of GRB host galaxies possibly be reproduced 
by the observational bias against dusty host galaxies.  

In this section, we have considered the possibility 
that the observational bias against dusty host galaxies 
is causing the systematic faintness of GRB host galaxies. 
However, we note that the SN observations possibly suffer 
more from dust extinction than the optical GRB afterglow observations. 
In that case, the observational bias against dusty host galaxies 
would make the SN host galaxy sample fainter than the GRB host galaxies, 
which is opposite to the observational result. 
Our simulation does not contradict with the observations
even if the SN observations suffer from dust extinction, 
because the simulated luminosity PDF with 
$E_{B-V, {\rm crit}} \geq 0.1$ agrees with the SN host observations 
reasonably well ($\Pks > 0.2$) for the Model 0.

\subsubsection{Fraction of Optically Dark GRBs}
\label{sec:darkfrac}
The physical nature of optically dark GRBs is also an important issue. 
A recent study by \citet{Cenko:09a} reported that the fraction 
of optically dark GRBs in the entire observed GRB sample is $\sim 50$\%, 
while \citet{Zheng:09a} found that the fraction is $\sim 10 - 20$\%. 
Note that these fractions are based on the entire GRB sample 
that span a wide redshift range, because it is 
difficult to constrain the redshift of optically dark GRBs. 

\begin{figure}
\includegraphics[scale=0.8]{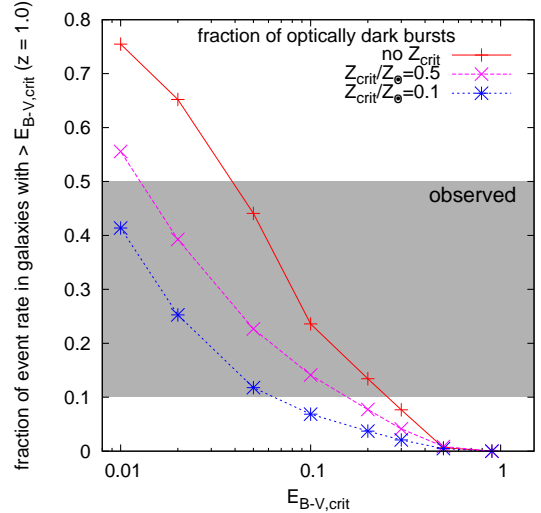}
\caption {
Fraction of optically dark GRBs at $z=1.0$ in our simulation 
(i.e., the fraction of GRB events in galaxies with $E_{B-V}>E_{B-V, {\rm crit}}$) 
as a function of $E_{B-V, {\rm crit}}$.  
The results for three different values of $\Zcrit$ are shown.  
The observed range of fractions for the GRBs at all redshifts 
\citep{Cenko:09a,Zheng:09a} is indicated with the shade.
We use Model 0 in this figure. 
}
\label{fig:darkfrac}
\end{figure}

In Figure~\ref{fig:darkfrac}, we show the fraction of GRB event rate 
in galaxies with $E_{B-V} > E_{B-V, {\rm crit}}$ at $z=1.0$ in our simulation 
for three different values of $\Zcrit$. 
Assuming that a GRB at $z=1.0$ becomes optically dark 
when the host galaxy has $E_{B-V} \geq E_{B-V, {\rm crit}} = 0.3$ 
\citep[$A_V\gtrsim 1,$][]{Perley:09a}, 
our simulation predicts that the fraction of optically dark GRBs is 
$\lesssim$ 10\% for all cases of $\Zcrit$. 
For the no-$\Zcrit$ case ({\it red solid} line), the dark fraction is 
close to 10\% for $E_{B-V, {\rm crit}} = 0.3$, and is almost consistent 
with the observational result of  \citet{Zheng:09a}. 
However, if $\Zcrit < 0.5\,\Zsun$ (as suggested in \S\,\ref{sec:result}), 
our simulation predicts that the fraction of optically dark GRBs 
is $< 5$\% for $E_{B-V, {\rm crit}} = 0.3$, which is much smaller than the observed fraction. 

The small fraction of optically dark GRBs in our simulation 
compared to the observations can be explained in two different ways. 
The first possibility is that 
the optically dark GRBs mainly suffer from 
the absorption in the optical wavelength due to dust
associated with small scale structures, 
not from the galaxy scale dust as modeled in \S\,\ref{sec:modelEbv}. 
The second possibility is that 
many of the observed optically dark GRBs 
originate at different redshifts other than $z\sim 1.0$, 
although the redshift distribution of GRBs with known redshifts 
that are typically optically non-dark, peaks at $z\sim 1.0$. 
The fractional contribution of dusty galaxies 
to the cosmic SFR would be different at different redshifts. 
Furthermore, the afterglow of a GRB at $z\gtrsim 6$ 
would be attenuated by neutral intergalactic medium 
in the observer-frame optical bands. 

It is unclear which of the above two scenarios might be correct at this point. 
Some observational studies \citep{Castro-Tirado:07a,Perley:09a,Holland:10a} 
find that the host galaxies of optically dark GRBs have bluer colors 
than those expected from the dust reddening estimated from 
the afterglow spectra. 
On the other hand, the host galaxies that is globally dusty are also found 
for several optically dark GRBs \citep{Levesque:10d, Hashimoto:10a}.

\subsection{Comparison with Previous Works}
\label{sec:previous}
Several studies have been performed to interpret 
the observed luminosity PDFs of GRB and SN host galaxies. 
In this section we compare our study with previous results. 

\citet{Wolf:07a} is one of the important studies in this context.
They reproduced the luminosity PDFs of the observed GRB and SN 
host galaxies based on the empirical relations 
of galaxy properties at $z \sim 0.7$, such as the luminosity-metallicity relationship. 
In their study, the model in which the GRBs 
occur in high metallicity regions ($Z\sim \Zsun$) reproduces 
the luminosity PDF of GRB host galaxy the best, 
and the models which requires GRBs to occur 
only in sub-solar metallicity regions do not reproduce the observations. 
They assumed a power-law relation for the 
luminosity-metallicity relation of galaxies following \citet{Kobulnicky:04a}. 
However this relationship is not well constrained by current observations
at redshifts of our concern, and in the work of \citet{Kobulnicky:04a}, 
different fitting methods gave different set 
of best fit parameters for a set of observed data, 
suggesting large uncertainties in the fit. 
We also note that the results probably depend on the assumed dispersion 
around the power-law luminosity-metallicity relation. 

\citet{Kocevski:09a} studied the stellar mass PDF of GRB host galaxies 
in a similar way to that of \citet{Wolf:07a} and argued for $\Zcrit > 0.5\,\Zsun$.
The mass-metallicity relationship of galaxies are better understood 
than the luminosity-metallicity relation \citep{Tremonti:04a,Savaglio:05a}. 
However, it is difficult to understand the selection effects 
of the currently available samples of GRB host galaxies 
with known $\Mstar$. 
\citet{Kocevski:09a} discussed the observed samples 
of GRB host galaxies with known $\Mstar$ collected 
by \citet{Castro-Ceron:08a} and \citet{Savaglio:09a}. 
Both of the samples were collected 
from publicly available data of GRB host observations. 
\citet{Castro-Ceron:08a} collected all GRB host galaxies 
with rest-frame $K$-band data including upper limits, 
while \citet{Savaglio:09a} selected their sample 
requiring multi-band detection of the host galaxies. 
The observations of GRB host galaxies in the samples 
are originally done by various groups with differing instrument, 
and hence the selection effects in the samples are hardly understood.
We also note that the constraints on the stellar masses 
for some of the GRB host galaxies in their sample 
are not accurate enough to quantitatively discuss 
the low-metallicity preference. 

The approaches of \citet{Wolf:07a} and \citet{Kocevski:09a} 
are entirely based on the empirical relations, 
and are fundamentally different from our approach 
based on numerical simulations. 
As discussed above, one difficulty in studying 
the luminosity PDFs of GRB/SN host galaxies is 
that properties of galaxies are not observationally well understood 
at redshifts where the observed sample is available. 
Numerical simulation approach is powerful in this context, 
because it gives properties of galaxies 
such as luminosity, metallicity, and SFR 
without assuming uncertain relations between them. 

Some numerical studies have investigated 
the luminosity of GRB host galaxies \citep{Nuza:07a,Lapi:08a,Campisi:09a}. 
However, quantitative comparisons of the predicted 
luminosity PDF of GRB host galaxies and the observed luminosity PDFs 
have not been performed in these studies. 
\citet{Nuza:07a} did not reproduce the luminosity PDFs of both 
GRB and SN hosts, 
presumably owing to the small box size of their simulation (10$h^{-1}$\,Mpc). 
\citet{Lapi:08a} and \citet{Campisi:09a} have shown that 
the fainter galaxies have systematically lower metallicities in their models. 
However, in \citet{Lapi:08a} and \citet{Campisi:09a}, 
the luminosity PDFs are calculated simply by selecting galaxies 
that contain young low-metallicity stars without taking the difference 
of GRB rate among galaxies into account, 
therefore the predicted luminosity PDFs cannot be 
directly compared with the observations of GRB host galaxies. 

In this context, our present work is unique in the sense 
that it can quantitatively reproduce the observations of 
both GRB and SN host galaxies using a self-consistent 
cosmological hydrodynamic simulations of galaxy formation. 
This has been enabled by applying a widely used GRB rate model 
to our cosmological simulations with large box sizes, 
which can properly compute the properties of galaxies at $z\sim 1$.

\section{Conclusions}
\label{sec:conclusion}

Using cosmological SPH simulations, 
we have examined the relation between the metallicity dependence of GRBs 
and the luminosity PDF of GRB host galaxies. 
Our results suggest that the observed difference in the UV luminosity PDFs 
of GRB/SN host galaxies can be explained 
by the low-metallicity preference of GRBs. 

We find that in our simulation, the luminosity PDF of GRB host galaxies 
agree with the observed one when $\Zcrit < 0.5\,\Zsun$, 
while the observed luminosity PDF of SN host galaxies
can be reproduced without a metallicity dependence of SN rate.  
The suggested value of $\Zcrit$ for the GRBs by our study is consistent 
with the suggestion from stellar evolution models 
\citep{Yoon:05a,Woosley:06a},  
contrary to the results of previous studies \citep{Wolf:07a,Kocevski:09a}.
Though our simulation suffers from 
the seeming overprediction of the UV luminosity function 
at $z=1$ at the bright end, 
we have explicitly demonstrated that the quantitative agreement 
between the simulated and observed luminosity PDFs is not a product 
of this seeming overestimate of the UV-bright galaxies in our simulation.  
However, we should be cautious about our results when 
$\Zcrit < 0.1\,\Zsun$, since they may be affected by 
the resolution limit of our simulation. 

We have also discussed the effect of observational bias 
against the host galaxies of optically dark GRB on the luminosity PDFs. 
The dust bias causes a shift of luminosity PDFs toward fainter magnitudes, 
similarly to the effect of the low-metallicity preference of GRBs. 
However, to obtain an acceptable fit to the observed luminosity PDF 
of GRB host galaxies, we had to assume that an optical afterglow 
becomes significantly faint when $E_{B-V}\sim 0.03$ in the host galaxy.
Such a small extinction is unlikely to cause a significant effect 
for the optical follow-up observations of GRBs, 
and therefore it would be difficult to explain the observed luminosity PDFs 
of GRB and SN host galaxies only by the overall dust bias of the host galaxies. 
Nevertheless it is possible that our simplified model of dust extinction 
is not sufficient to reproduce the dust extinction of GRB optical afterglows. 

Assuming that a GRB becomes optically dark 
when the extinction of its host galaxy is $E_{B-V}\gtrsim 0.3$ 
as suggested by the observation of \citet{Perley:09a}, 
our simulation predicts that the fraction of optically dark GRBs 
at $z=1.0$ is $< 5$\% for $\Zcrit \leq 0.5\,\Zsun$. 
This fraction is significantly smaller than what is suggested 
by the observations of GRBs at all redshifts \citep{Perley:09a,Zheng:09a}. 
This is probably because we did not consider 
the dust extinction effect on small scales at the GRB site. 
It is also possible that the GRBs at redshifts other than $z\sim 1$ 
are important in determining the fraction. 
We plan to address the effect of small scale dust 
in the future using higher resolution simulations.

\acknowledgements
This work was supported by the Grant-in-Aid for the Global COE Program 
"The Next Generation of Physics, Spun from Universality and Emergence" 
from the Ministry of Education, Culture, Sports, Science 
and Technology (MEXT) of Japan.
This work was also supported in part by the NSF grant AST-0807491, 
National Aeronautics and Space Administration under Grant/Cooperative 
Agreement No. NNX08AE57A issued by the Nevada NASA EPSCoR program, 
the President's Infrastructure Award from UNLV, 
and by the NSF through the TeraGrid resources provided by the Texas Advanced Computing Center.
Some numerical simulations and analyses have also been performed on the UNLV Cosmology Cluster. 
KN and YN are grateful for the hospitality of the Institute for the Physics and 
Mathematics of the Universe (IPMU), University of Tokyo, where part of this work was done.

\bibliography{reference_list}

\end{document}